\documentclass[aps,prb,twocolumn]{revtex4-1}
\usepackage{amssymb}
\usepackage{graphicx}

\begin{document}
\author{Kavita Mehlawat and Yogesh Singh }
\affiliation{$^1$Department of Physical Sciences, Indian Institute of Science Education and Research (IISER) Mohali, Knowledge City, Sector 81, Mohali 140306, India.}

\date{\today}

\title{Quantum Spin Liquid in a depleted triangular lattice Iridate K$_x$Ir$_y$O$_2$}

\begin{abstract}
We report discovery of a new iridate family K$_x$Ir$_y$O$_2$ with depleted triangular lattice planes made up of edge sharing IrO$_6$ octahedra separated by K planes.  Such a material interpolates between the triangular and honeycomb lattices and is a new playground for Kitaev physics.  The materials are Mott insulators with $y = 1 - x/4$.  Physical property measurements for the $x \approx 0.85$ material are reported.  Using magnetic susceptibility $\chi$ versus temperature $T$ measurements we find $S_{eff} = 1/2$ moments interacting strongly with a Weiss temperature $\theta \approx - 180$~K and no magnetic order or spin freezing down to $T = 1.8$~K\@.  Heat capacity shows a broad maximum around $30$~K which is insensitive to magnetic fields and a $T$-linear low temperature behaviour with $\gamma \sim 10$~mJ/mol~K$^2$.  These results are consistent with a gapless QSL state in K$_{0.85}$Ir$_{0.79}$O$_2$. 
\end{abstract}

\maketitle  

Geometrical frustration has been the preferred recipe to realize the ellusive Quantum Spin Liquid (QSL) state in local moment magnets \cite{Mila2000, Balents2010}.  Recently however, dynamic frustration in systems with strongly anisotropic magnetic exchange has been shown to be another avenue to explore the QSL state.  Kitaev proposed a honeycomb lattice model of spins $S = 1/2$ interacting via bond dependent Ising interactions and showed that it was exactly solvable in the Majorana representation \cite{Kitaev2006}.  This is an example of a model with dynamic frustration leading to a QSL ground state \cite{Kitaev2006}.  Jackeli and Khaliullin gave a recipe to engineer such Kitaev interactions in real materials via strong spin-orbit coupling \cite{Jackeli2009}.  It was however, realized that additional interactions could not be avoided in real materials due to deviations from the ideal structure needed for perfect cancellation of Heisenberg exchange, and also due to direct $d$-$d$ exchange \cite{Chaloupka2010}.  Subsequently, more general Hamiltonians have been considered which include further neighbour Heisenberg, Kitaev, and/or off-diagonal Gamma exchange interactions \cite{Winter2017}. 

The quasi-two dimensional layered honeycomb materials Na$_{2}$IrO$_{3}$, $\alpha$-Li$_2$IrO$_3$, and $\alpha$--RuCl$_3$ and the three dimensional iridates $\gamma$--Li$_2$IrO$_3$ and $\beta$--Li$_2$IrO$_3$ have garnered a lot of attention as prime candidate materials to host Kitaev physics \cite{Winter2017, Trebst2017, Jackeli2009, Chaloupka2010, Singh2010, Liu2011, Kimchi2011, Choi2012, Ye2012, Singh2012, Gretarsson2012, Comin2012, Chaloupka2013, Gretarsson2013, Foyevtsova2013, Katukuri2014, Yamaji2014, Sizyuk2014}.  Although there is growing evidence of dominant Kitaev interactions from $ab ~initio$ estimations of the exchange parameters \cite{Foyevtsova2013, Katukuri2014,Yamaji2014} as well as from direct evidence of dominant bond directional exchange interactions in Na$_{2}$IrO$ _{3}$\cite{Chun2015}, these materials were unfortunately found to be magnetically ordered at low temperatures \cite{Singh2010, Choi2012, Ye2012, Singh2012}.  The intense search for Kitaev materials which do not show conventional magnetic order has borne fruit recently with the discovery of two new honeycomb lattice iridate materials H$_3$LiIr$_2$O$_6$ ~\cite{Kitagawa2018} and Cu$_2$IrO$_3$~ \cite{Abramchuk2017}.   Experimental results on both materials are consistent with a quantum spin liquid state \cite{Kitagawa2018, Choi2019} although their connection with the Kitaev QSL is still a matter for future investigation.  

The application of the Kitaev-Heisenberg model has been extended to other two- (2D) and three-dimensional (3D) lattices recently.  These lattices include the 2D triangular, honeycomb, and Kagome lattices, and the 3D FCC, Pyrochlore, and hyperkagome lattices \cite{Kimchi2014,Rousochatzakis}.  Unconventional magnetic states including a vortex crystal state as well as quantum paramagnetic states exist in the generic phase diagram of these lattices \cite{Kimchi2014,Rousochatzakis}.  While there are iridate material realizations of most of the lattices studied above, a material with a triangular iridium lattice isn't currently available.  The layered materials $A_x$CoO$_2$ \cite{Huang2004,Jansen1974,Hironaka2017,Nakamura1996} and $A_x$RhO$_2$ ($A = ~$Na, K) \cite{Shibasaki2010,Zhang2013} indeed adopt a structure made up of layers of edge-sharing CoO$_6$ or RhO$_6$ octahedra on a triangular lattice.  However, because of weaker spin-orbit coupling these materials are found to be metallic for a large range of $x$.
 

In this work we report crystal growth of a new family of spin-orbit Mott insulators K$_x$Ir$_y$O$_2$ ($x, y \leq 1$).  The structure of these materials has been reported recently using a combined single crystal diffraction and DFT calculation study \cite{Johnson2019}.  The structure is built up of layers of edge-sharing IrO$_6$ octahedra arranged on a triangular lattice, and the K binds the layers together.  The potassium content $x$ controls the charge in the triangular layers similar to the K$_x$CoO$_2$ type materials mentioned above.  The important difference is that while K$_x$CoO$_2$ materials are metallic for a range of $x$ values, the K$_x$Ir$_y$O$_2$ materials stay Mott insulating by creating Ir vacancies so that $y = 1-x/4$~ \cite{Johnson2019}.   As $x$ increases from $x = 0$, Ir vacancies are created to maintain charge neutrality keeping the Ir$^{4+}$ valance state.  These vacancies are distributed randomly on the triangular lattice.  However, as $x$ increases above some critical value the vacancies form an ordered structure and occupy voids in a honeycomb Ir lattice. Thus, when $x < x_c \approx 0.8$, the materials adopt a triangular structure similar to the cobaltates K$_x$CoO$_2$ whereas for $x > x_c$ the Ir vacancies adopt an ordered structure such that they are located in the voids of an Ir honeycomb lattice similar to Na$_2$IrO$_3$.  In fact for $x = 4/3$ we obtain the hypothetical material K$_2$IrO$_3$ with a structure made up of honeycomb layers of Ir$^{4+}$ ions separated by K layers.  Therefore K$_x$Ir$_y$O$_2$ materials present a unique opportunity of studying Kitaev physics in layered iridates which interpolate between the triangular lattice to the honeycomb lattice.  The structure can be tuned in principle between these two limits by controlling the potassium content $x$.  

\begin{figure}[t]   
\includegraphics[width= 2 in]{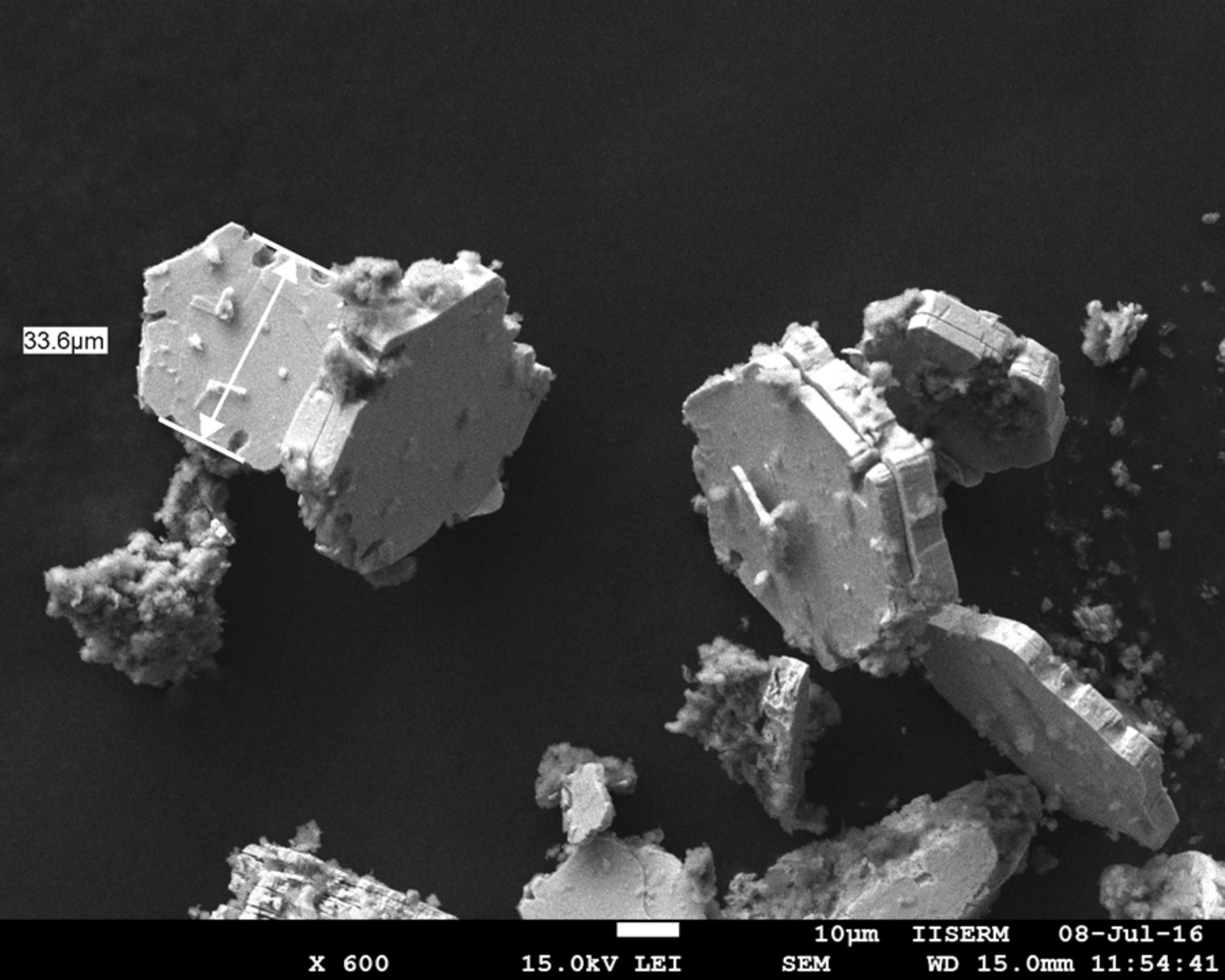}
\includegraphics[width= 2.5 in]{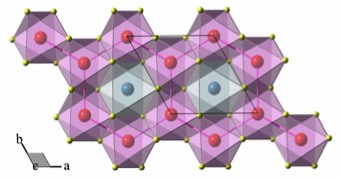}    
\caption{(Color online) (top) Scanning electron images of several K$_x$Ir$_y$O$_2$ crystals showing hexagonal morphology which reflects their underlying crystal structure.  (bottom) The $ab$-plane view of honeycomb layers formed by edge-sharing IrO$_6$ octahedra (pink).  The red spheres are the Ir, the yellow spheres are the oxygen.  The voids in the honeycomb lattice are partially occupied by Ir$^{4+}$ ions (blue).    
\label{Fig-structure}}
\end{figure}

Single crystals of K$_x$Ir$_{1-x/4}$O$_2$ were grown from metallic K (99.995\%, Alfa Aesar) and Ir powder (99.95\%, Alfa Aesar) taken in the ratio $2:1$ and placed inside Al$_2$O$_3$ crucibles in air, similar to the method  of separated euducts used recently for crystal growth of $\alpha$-Li$_2$IrO$_3$~\cite{Freund2016}.  Small broken pieces of Al$_2$O$_3$ crucibles were placed inside the growth crucible.  This was necessary to allow nucleation of crystals at sharp edges above the floor of the growth crucible.  The growth crucible with Ir metal powder on the floor and K metal on top of the broken Al$_2$O$_3$ pieces, was heated to $1070~^o$C in $12$~hrs and left there for $70$~hrs before cooling to room temperature in $12$~hrs.  We were able to grow crystals with $x = 0.61$ with a triangular structure and $x = 0.85$ with a honeycomb structure.  The structural details of both kinds of crystals have been reported recently \cite{Johnson2019}.  The $x = 0.6$ crystals were too small in size and not enough in number for physical property measurements.  In this work we therefore report physical property measurements on a collection of crystals with $x = 0.85(5)$.  A scanning electron microscope image of the $x = 0.85$ crystals is shown in Fig.~\ref{Fig-structure} and clearly reveals the underlying honeycomb symmetry and the layered nature of the structure.

Our main results are that temperature dependent magnetic susceptibility measurements confirm local moment magnetism (Mott insulator) with effective spins $S_{eff} = 1/2$ interacting strongly $\theta = -180$~K\@.  Both magnetic susceptibility and heat capacity show that these interacting moments do not undergo a transition to a long ranged magnetically ordered state or to a spin-glass state down to $T = 1.8$~K\@.  Heat capacity shows a broad maximum around $30$~K which is insensitive to magnetic fields and a $T$-linear low temperature behaviour with $\gamma \sim 10$~mJ/mol~K$^2$.  These results are consistent with a gapless QSL state in K$_{0.85}$Ir$_{0.79}$O$_2$. \\  

\noindent
\emph{Structure:} The structure of K$_{0.85}$Ir$_{0.79}$O$_2$ was arrived at using a combination of single crystal diffraction and density functional theory (DFT) calculations \cite{Johnson2019}.  These crystals adopt a honeycomb structure with space group P$6_{3}22$ (no. $182$) and unit cell parameters $a = b = 5.2823(2)~$\AA~ and $c = 13.5437(7)~$\AA.   The structure is built up of edge-sharing IrO$_6$ octahedra forming a honeycomb lattice.  The voids in the honeycomb lattice are partially occupied by Ir.  These honeycomb Ir layers are separated by K layers \cite{Johnson2019}.

The distance between the Ir honeycomb layers in K$_{0.85}$Ir$_{0.79}$O$_2$ is found to be $6.8$~\AA.  This distance is $\approx 30\%$ larger than the distance ($5.3$~\AA) between Ir honeycomb layers in Na$_2$IrO$_3$.  This suggests that enhanced quantum fluctuations due to increased low-dimensionality can be expected.  The Ir honeycomb in K$_{0.85}$Ir$_{0.79}$O$_2$ is found to be perfect with all Ir-Ir distances equal to $3.047$~\AA.  Additionally, the IrO$_6$ octahedra are also nearly ideal with Ir-O distances ranging from $2.01$~\AA to $2.02$~\AA and Ir-O-Ir angle equal to $93.65~^\circ$.  

\begin{figure}[t]   
\includegraphics[width= 3 in]{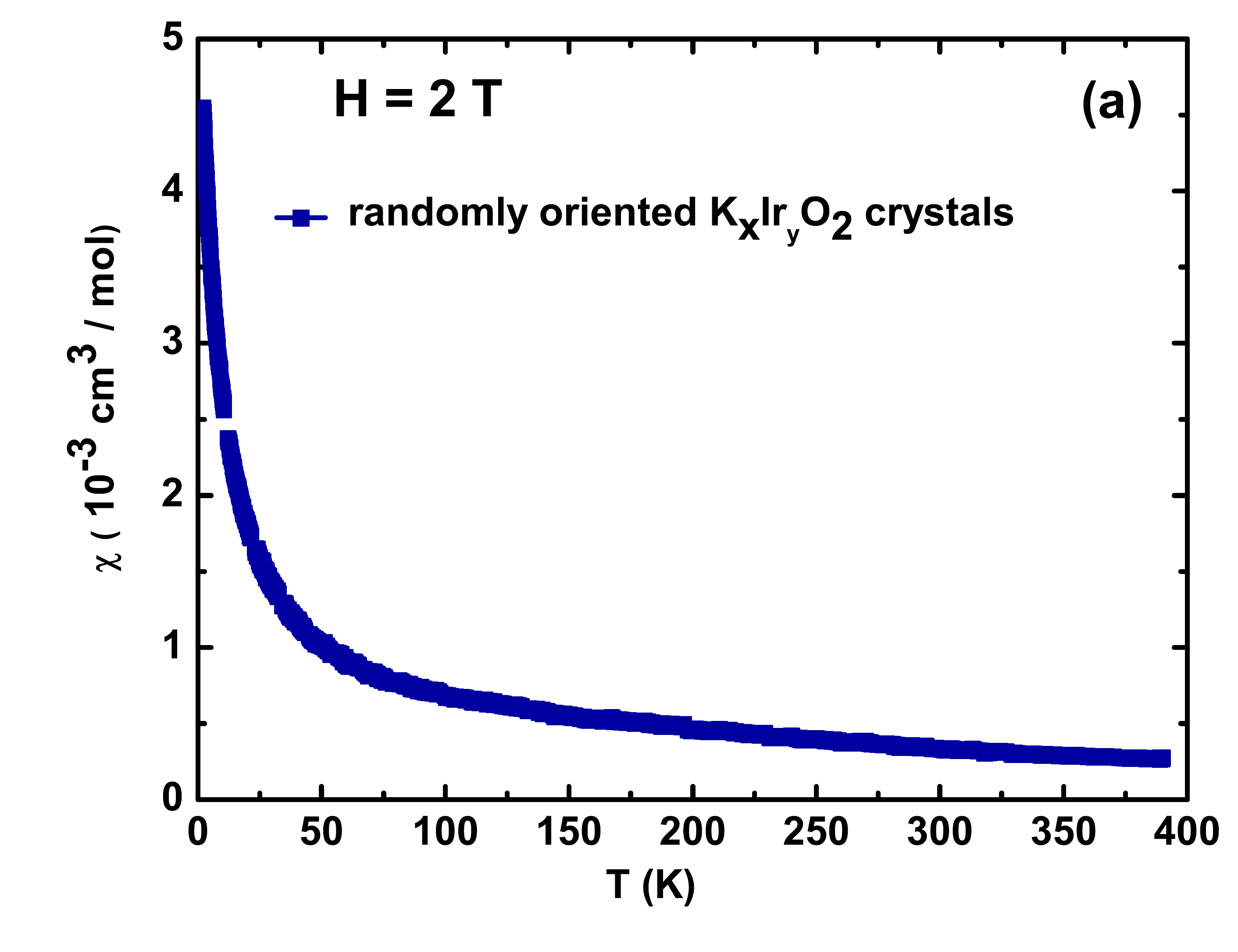}
\includegraphics[width= 3 in]{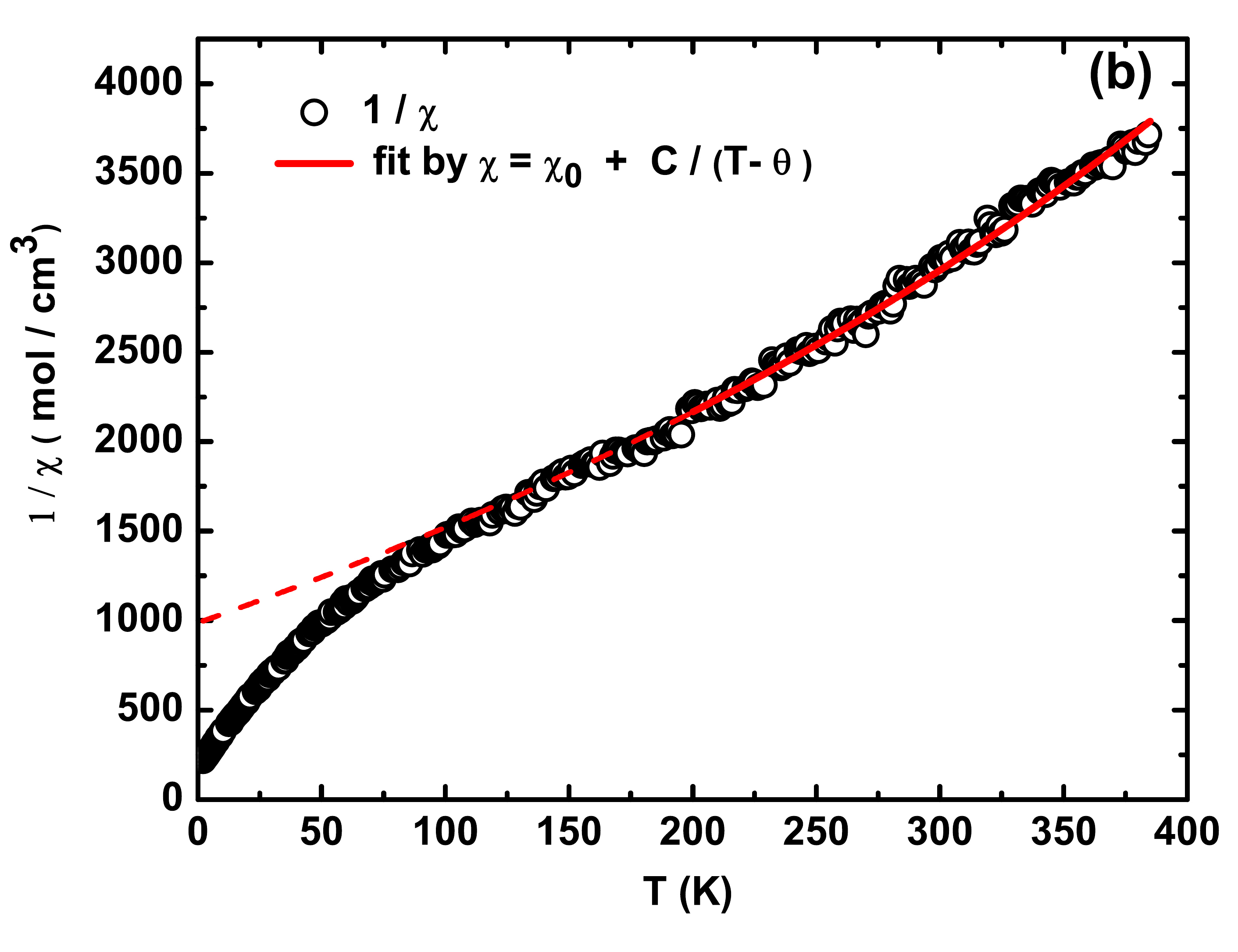}    
\caption{(Color online) (Top) Magnetic susceptibility $\chi$ versus temperature $T$ for randomly oriented K$_x$Ir$_y$O$_2$ crystals measured between $T = 1.8$~K and $400$~K in a magnetic field of $H = 2$~T\@.  (Bottom)  The $1/\chi(T)$ data for K$_x$Ir$_y$O$_2$ crystals.  The solid curve through the data is a fit to the high temperature data by a Curie-Weiss expression.  The parameters obtained from the fit are shown (see text for details).  
\label{Fig-chi}}
\end{figure} 

\noindent
\emph{Magnetic Susceptibility:}
Figure~\ref{Fig-chi} shows the magnetic susceptibility $\chi$ versus temperature $T$ data measured in a magnetic field of $H = 2$~T for a collection of randomly oriented K$_{0.85}$Ir$_{0.79}$O$_2$ crystals.  The $\chi$ shows a strong $T$ dependent behaviour consistent with localized moment magnetism.  Together with the insulating transport \cite{Johnson2019} this confirms the Mott insulating state of these crystals.  There is no signature of any long range magnetic order down to $T = 1.8$~K\@.  A low field measurement (not shown) at $H = 100$~Oe did not show any cusp in $\chi(T)$ which suggests absence of spin freezing. 
 
The $H = 2$~T data are plotted as $1/\chi(T)$ in the lower panel of Fig.~\ref{Fig-chi}.  The data above $T \approx 200$~K were fit by the expression $\chi(T) = \chi_0+{C\over T-\theta}$, with $\chi_0$ , $C$, and $\theta$ as fit parameters.  Here $\chi_0$ is a $T$ independent term, $C$ is the Curie constant, and $\theta$ is the Weiss temperature.  The fit gave the values $\chi_0 = -1.1(7) \times 10^{-4}$~cm$^3$/mol, $C = 0.391(3)$~cm$^3$~K/Ir~mol, and $\theta = -180(9)$~K\@.  The large diamagnetic value for $\chi_0$ is most likely in part due to contributions from the sample holder.  The value of the Curie constant $C = 0.391$~cm$^3$~K/Ir~mol is close to what is expected ($C = 0.375$~cm$^3$~K/Ir~mol ) for $S = 1/2$ moments with an electronic $g$-factor $g = 2$.  The value of $\theta$, which gives the overall scale of the magnetic interactions, comes out to be large and negative suggesting predominance of antiferromagnetic interactions.  The absence of any static magnetic order down to $T = 1.8$~K indicates a strongly frustrated material. \\

\begin{figure}[t]   
\includegraphics[width= 2.75 in]{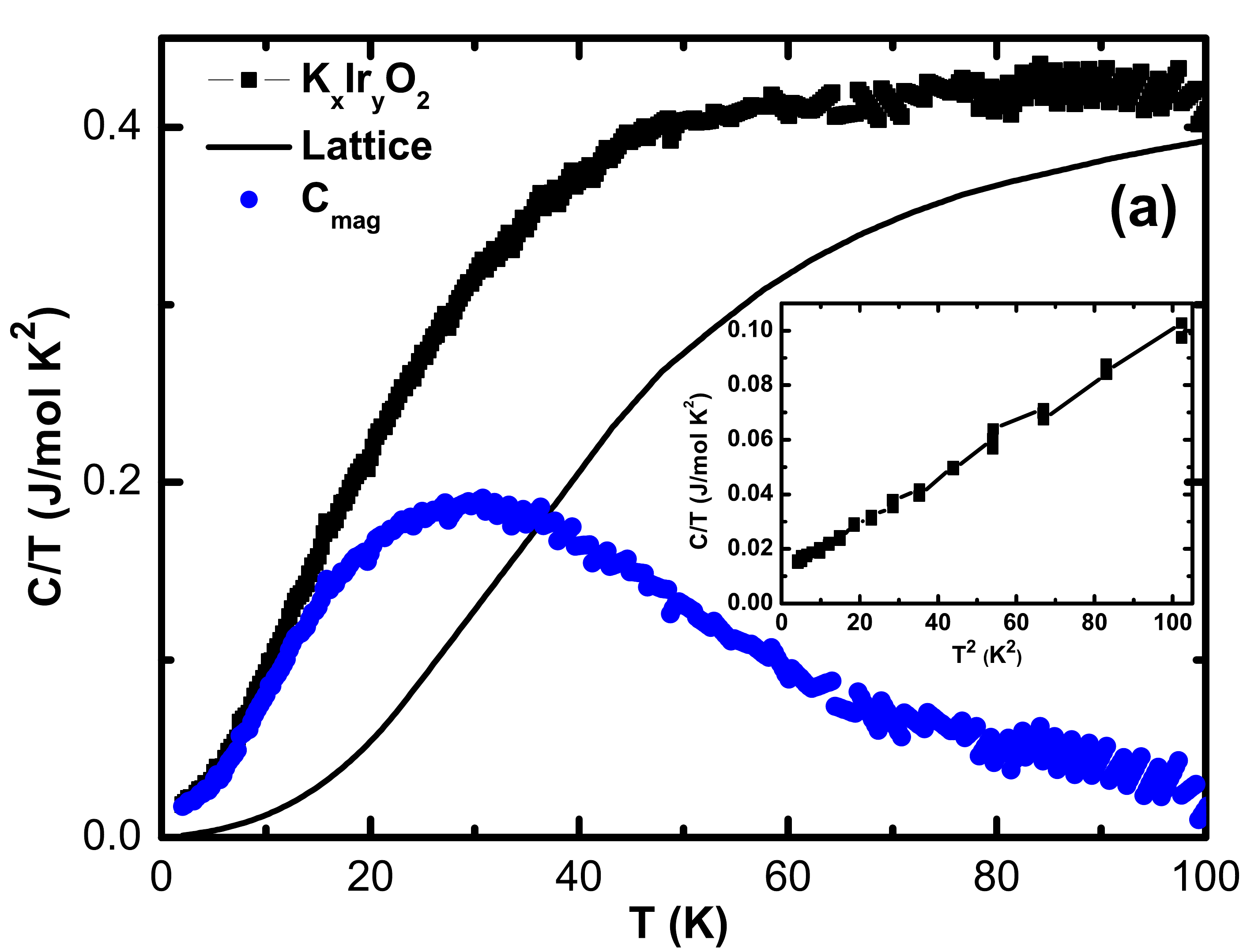}
\includegraphics[width= 2.75 in]{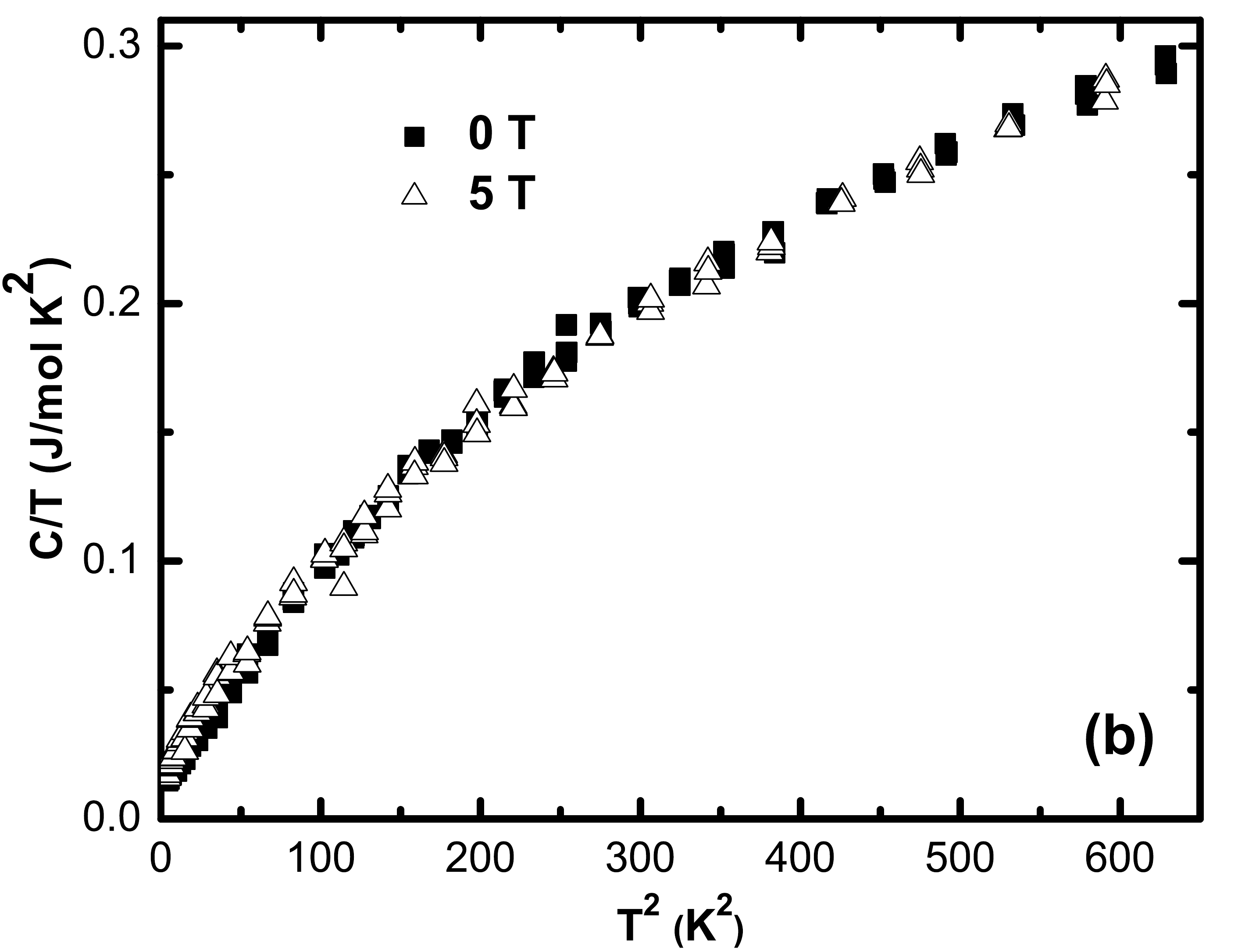}
\caption{(Color online) (a) Heat capacity $C$ for a collection of K$_{0.85}$Ir$_{0.79}$O$_2$ crystals, the approximate lattice contribution, and the inferred magnetic contribution $C_{\rm mag}$ versus $T$.  The inset shows the $C/T$ versus $T^2$ data at low temperatures.  (b) The $C/T$ data at magnetic fields $H = 0$ and $5$~T\@.  The inset shows the power-law dependence of $C(T)$ at low temperatures.  The exponent of the power-law changes in a field (see text for details).  
\label{Fig-Cp}}
\end{figure} 

\noindent
\emph{Heat Capacity:}  Figure~\ref{Fig-Cp}~(a) show the heat capacity divided by temperature $C/T$ versus $T$ data for a collection of crystals ($\sim 8$~mg) of K$_{0.85}$Ir$_{0.79}$O$_2$ between $T = 1.8$~K and $100$~K\@.  There is no sharp anomaly which indicates absence of any conventional magnetic transition despite the strong exchange interactions indicated by $\theta = -180$~K found from magnetic measurements above.  To estimate the magnetic contribution to the heat capacity of K$_{0.85}$Ir$_{0.79}$O$_2$, we use the heat capacity of Na$_2$SnO$_3$ as an approximate lattice contribution for K$_{0.85}$Ir$_{0.79}$O$_2$.  First Na$_2$SnO$_3$ is written as Na(Na$_{1/3}$Sn$_{2/3}$)O$_2$ and the heat capacity of Na(Na$_{1/3}$Sn$_{2/3}$)O$_2$ obtained by rescaling the Na$_2$SnO$_3$ data for the mass difference. The Debye temperature for Na$_2$SnO$_3$ is about $\theta_D = 650$~K\@.  At low temperatures ($\theta_D/10$) the contribution to the lattice heat capacity will be mostly from long wave-length phonons.  Since both K$_{0.85}$Ir$_{0.79}$O$_2$ and Na$_2$SnO$_3$ structures are built up of similar honeycomb layers stacked along the $c$-axis, we believe that Na$_2$SnO$_3$ can be used to get an approximate lattice heat capacity contribution, specially at low temperature.  

The magnetic contribution to the heat capacity $C_{\rm mag}$ obtained by subtracting the heat capacity of Na(Na$_{1/3}$Sn$_{2/3}$)O$_2$ from the heat capacity of K$_{0.85}$Ir$_{0.79}$O$_2$ is also shown in Fig.~\ref{Fig-Cp}~(a) plotted as $C_{\rm mag}/T$ versus $T$.  The $C_{\rm mag}/T$ shows a broad anomaly peaked at $30$~K with an extended tail to higher temperatures.  The entropy obtained by integrating the $C_{\rm mag}/T$ vs $T$ data from $1.8$~K to $100$~K is close to $50\%$Rln$2$ which is close to half of the value expected for $S = 1/2$.  This suggests that the rest of the entropy might be recovered either at lower temperatures through some magnetic ordering, or the entropy may be recovered at high temperatures upto $T = |\theta| \sim 200$~K\@.  For a Mott insulator, the lowest temperature $C(T)$ data follows an unconventional $\gamma T + \beta T^3$ dependence as shown in the Fig.~\ref{Fig-Cp}~(a) inset.  We obtain a large value $\gamma = 10$~mJ/mol~K$^2$ of the $T$-linear contribution to $C(T)$ suggesting gapless excitations.   

There is no significant change in $C(T)$ on the application of a magnetic field as can be seen from Fig.~\ref{Fig-Cp}~(b) where the $C(T)$ data measured in a field $H = 5$~T is plotted along with the zero field data.  The broad anomaly in $C_{mag}$ and absence of a magnetic field dependence is similar to the behaviour seen in several quantum spin liquid materials like.

\noindent
\emph{Summary and Discussion:} We report discovery and crystal growth of a new honeycomb lattice iridate family K$_x$Ir$_y$O$_2$ and report physical properties on the specific material K$_{0.85}$Ir$_{0.79}$O$_2$.  To make a comparison with Na$_2$IrO$_3$, we re-write it as Na(Na$_{1/3}$Ir$_{2/3}$)O$_2$.  The elements (Na,Ir) in the parenthesis form a triangular lattice with the Na sitting in the voids of honeycombs formed by the Ir.   These triangular layers are separated by Na layers.  We can re-write our material as K$_{0.85}$(Ir$_{0.39/3}$Ir$_{2/3}$)O$_2$.  This indicates that the honeycomb lattice formed by edge-sharing IrO$_6$ is fully occupied and about $61\%$ of the voids in the honeycomb lattice are vacant, while $39\%$ of the voids are occupied by Ir.   This is substantiated by single crystal XRD and DFT calculations \cite{Johnson2019}.  K$_{0.85}$Ir$_{0.79}$O$_2$ can therefore be viewed either as a depleted triangular lattice or a stuffed honeycomb lattice made of edge-sharing IrO$_6$ octahedra.  The structure therefore presents a unique platform to study Kitaev and Kitaev-Heisenberg physics on a lattice which interpolates between the triangular and honeycomb lattices.  Phase diagrams of the KH model on both lattices have been worked out.  On the honeycomb lattice several magnetic states as well as quantum spin liquid states exist while on the triangular lattice unconventional magnetic orders including a vortex crystal lattice are expected.  Studies starting from either lattice and going towards the other limit are not available yet but can be expected to reveal further novel magnetic phases.  

The physical properties of K$_{0.85}$Ir$_{0.79}$O$_2$ reported here indeed reveal unconventional behaviours.  The temperature dependent resistance shows the insulating nature of the material.  Magnetic susceptibility $\chi(T)$ of randomly oriented crystals shows localized moment behaviour confirming the Mott insulating state in K$_{0.85}$Ir$_{0.79}$O$_2$.  A Curie-Weiss analysis of the $\chi(T)$ data gives a value $S_{\rm eff} = 1/2$ for the Ir moments and we also get a large negative Weiss temperature $\theta = -180$~K indicating strong antiferromagnetic exchange between the $S = 1/2$ moments.  Despite this there is no signature of any conventional (long-range order or spin-glass) magnetic state down to $T = 1.8$~K\@.  The magnetic contribution to the heat capacity $C_{\rm mag}$ shows a broad anomaly peaked at $T \approx 30$~K which is magnetic field insensitive and carries approximately $50\%$ of the entropy expected for $S = 1/2$ moments.  The low temperature $C(T)$ shows a linear in $T$ dependence with a $T = 0$ intercept $\gamma = 10$~mJ/mol~K$^2$.  This large value in an insulating material suggests the presence of gapless excitations of unconventional nature. 

All these results point to an unconventional magnetic ground state and are consistent with a gapless quantum spin liquid.  Future studies will be required to understand the origin of these unconventional behaviours.  The possible role of disorder (apparent in the stacking faults) in producing the power-law $C(T)$ at low temperatures also needs investigation.  It is clear that there would be deviations from a pure Kitaev Hamiltonian since the Weiss temperature is large and negative instead of ferromagnetic as for the original Kitaev model.  Additionally, for the isotropic nearest-neighbor Kitaev model a $C \propto T^2$ dependence is expected.  A $T$-linear dependence observed by us therefore also suggests possible deviations from the pure Kitaev model.  Microscopic magnetic measurements down to lower temperatures and theoretical estimates of the exchange interactions would be essential in future to place K$_{0.85}$Ir$_{0.79}$O$_2$ on the map of generalized Kitaev Hamiltonians \cite{Winter2017}.  Nevertheless, our results demonstrate that K$_{0.85}$Ir$_{0.79}$O$_2$ is a new platform for the exploration of Kitaev physics in a lattice which interplotes between the triangular and honeycomb lattices and is sure to spawn many future experimental and theoretical studies.\\
         
\noindent
\emph{Acknowledgments.--} We thank the X-ray, the SEM, and the SQUID magnetometer facilities at IISER Mohali.  KM acknowledges UGC-CSIR India for a fellowship.  YS acknowledges DST, India for support through Ramanujan Grant \#SR/S2/RJN-76/2010 and through DST grant \#SB/S2/CMP-001/2013.

\end{document}